\documentclass{osa-article}

\usepackage{amsmath, commath, relsize}
\usepackage[T1]{fontenc}

\newcommand{\dd}[1]{\mathrm{d}{#1}}
\newcommand{\va}[1]{\mathbf{#1}}
\newcommand{\sref}[1]{Sec.~\ref{#1}}
\graphicspath{{./figs/}}

\newcommand{\ie}{i.e.~}

\newcommand{\eg}{e.g.~}

\journal{oe}

\begin{document}

\title{Plasmons in ultra-thin gold slabs with quantum spill-out: Fourier modal method, perturbative approach, and analytical model}

\author{Alireza Taghizadeh,\authormark{1,2,*} and Thomas Garm Pedersen \authormark{1,2}}

\address{\authormark{1}Department of Materials and Production, Aalborg University, 9220 Aalborg {\O}st, Denmark\\
\authormark{2}Center for Nanostructured Graphene (CNG), 9220 Aalborg {\O}st, Denmark\\}

\email{\authormark{*}ata@nano.aau.dk} 



\begin{abstract}
We numerically study the effect of the quantum spill-out (QSO) on the plasmon mode indices of an ultra-thin metallic slab, using the Fourier modal method (FMM). To improve the convergence of the FMM results, a novel nonlinear coordinate transformation is suggested and employed. Furthermore, we present a perturbative approach for incorporating the effects of QSO on the plasmon mode indices, which agrees very well with the full numerical results. The perturbative approach also provides additional physical insight, and is used to derive analytical expressions for the mode indices using a simple model for the dielectric function. The analytical expressions reproduce the results obtained from the numerically-challenging spill-out problem with much less effort and may be used for understanding the effects of QSO on other plasmonic structures.
\end{abstract}

\section{Introduction}

Surface plasmon polaritons (SPPs) are transverse magnetic (TM) surface modes, which propagate at metal-dielectric interfaces \cite{Maier2007}. In the last two decades, there has been a tremendous interest in SPPs due to their ability to guide light in a volume much smaller than typical dielectric structures \cite{Takahara1997, Zia2004}. They have found numerous applications in solar cells, surface-enhanced Raman spectroscopy, lasers, sensors, etc \cite{Barnes2003, Schuller2010, Gramotnev2010, Zhang2012}. When a nm-thick metal slab is surrounded by dielectric media, the SPPs at the two metal-dielectric interfaces will strongly couple and form so-called short- and long-range SPPs \cite{Sarid1981}. In particular, the long-range SPPs have smaller propagation losses than the single interface SPPs \cite{Charbonneau2000}, which make them attractive for several applications such as nonlinear optics \cite{Berini2009} and opto-electronic devices \cite{Leosson2006}.

It is known that at the metal surface, the electron density has an exponential tail penetrating into the dielectric region due to the barrier tunneling \cite{Lang1970}. This quantum mechanical effect, which is typically referred to as quantum spill-out (QSO), has been shown to have significant impact on the optical properties of nano-scale metallic structures \cite{Zuloaga2010, Esteban2012, David2014, Zhu2016, Enok2018, Enok2019}. For a metallic slab, the QSO effect softens the abrupt changes of the dielectric function at the interfaces and leads to a position-dependent dielectric function across the slab. Therefore, numerical tools are required for calculating the plasmon mode indices when the QSO is included. This is in contrast to a classical metallic slab with spatially constant dielectric response, where the short- and long-range plasmon modes are obtained analytically. For a nm-thick metal slab, solving the QSO problem is numerically challenging due to the two vastly different length scales: the slab thickness ($\sim$1 nm) and the mode exponential decay in the dielectric regions ($\sim$1000 nm), particularly for the long-range SPPs. 
Recently, we have examined the effect of QSO on plasmons propagating in ultra-thin gold films \cite{Enok2018, Enok2019}, using a transfer matrix method. In this approach, the plasmon mode indices are obtained by finding the poles of the transfer matrix, relating the fields to the left and right of the slab \cite{Enok2018}. In the transfer matrix method, a staircase approximation should be employed for modeling the dielectric function in the vicinity of the interface (slicing the region into a large number of segments). To eliminate the discretization errors, the segment width should be as thin as 10$^{-4}$ nm \cite{Enok2019}, which makes the calculations cumbersome. Furthermore, the transfer matrix method requires reasonable initial guesses in order to find the poles efficiently, and one can only find a single plasmon at each run.


In the present work, we propose an alternative numerical approach for finding the plasmon mode indices of a thin metallic slab with QSO, based on the Fourier modal method (FMM). In this approach, all plasmon mode indices are obtained simultaneously by solving an eigenvalue problem in reciprocal (Fourier) space. To obtain acceptable numerical results with the FMM, we have implemented the fast factorization rule for TM polarization \cite{Li1996} and a nonlinear coordinate transformation. Inspired by Ref.~\cite{Hugonin2005}, we suggest a coordinate transformation, which drastically improves the convergence performance of the FMM for long-range SPPs.
Furthermore, since the QSO correction to the plasmon mode index is typically small, a perturbation technique is proposed, which reproduces the numerical results very accurately with much less effort. The perturbative approach requires only the classical solutions of the slab waveguide problem, which are obtained analytically. Moreover, using a simple model for the dielectric function, which mimics the dielectric function obtained from density-functional theory (DFT), analytical expressions are derived for the plasmon mode indices. We provide numerical results for thin gold slabs, and, by direct comparison, confirm the validity of the perturbative approach.
This paper is organized as follows. In \sref{sec:Theory}, we define the problem, and present its solution with or without QSO. In addition, we briefly review the FMM with a coordinate transformation and derive the perturbation expression for the plasmon mode index in this section.
In \sref{sec:Numerics}, the numerical results for thin gold slabs are provided, and the validity of our perturbative approach is demonstrated. Finally, a summary of the main findings is provided in \sref{sec:Conclusion}. A set of appendices provide checks of the validity of the perturbative approach, present the derivation of analytical expressions and explain the details of the coordinate transformation.

\section{Theoretical framework \label{sec:Theory}} 
\begin{figure}[h!]
	\centering\includegraphics[width=0.8\textwidth]{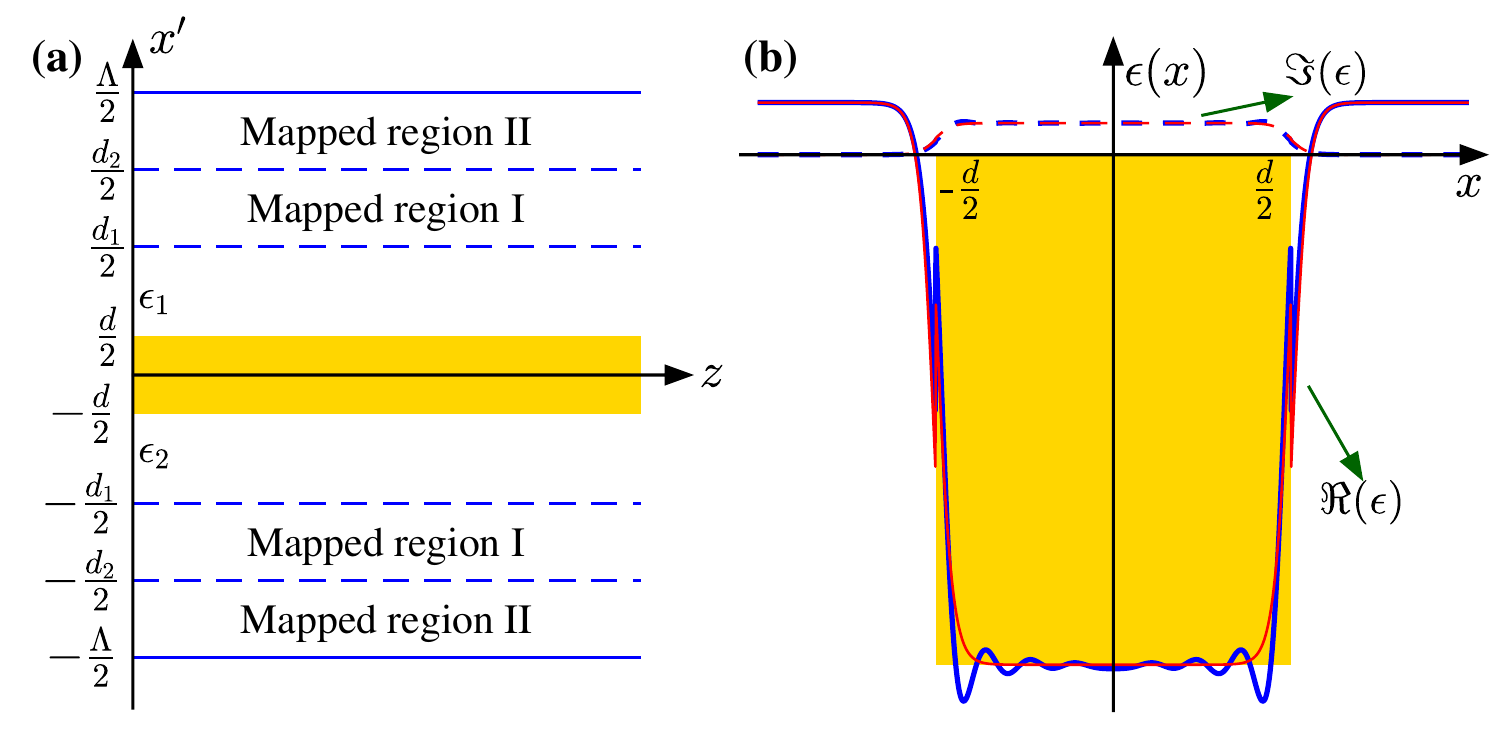}
	\caption{(a) Schematic of a slab waveguide with a thickness of $d$, with two mapped regions used for coordinate transformation $x=F(x')$, see the text. We choose the mapped regions to be far from the waveguide to avoid overlap with the regions affected by the QSO, \eg $d_1=d+20a$. (b) Real (solid) and imaginary (dashed) parts of the dielectric function $\epsilon(x)$ across the metal slab including the QSO ($d=2$ nm). The blue curve is a typical profile obtained from a DFT calculations (using a jellium model), whereas the red curve shows a fitted profile using Eq.~(\ref{eq:Tanh}) with $a=0.09$ nm.
	\label{fig:Schematic}}
\end{figure}

The plasmonic modes of a metallic slab waveguide, illustrated schematically in Fig.~\ref{fig:Schematic}(a), are obtained by finding the solutions to Maxwell's equations for TM polarization (p-polarization), written for the magnetic field $\va{H}$. The propagation direction is taken to be the $z$-direction, whereas the finite slab thickness is along the $x$-direction. The magnetic field is of the form $\va{H}=H_y(x)\exp(i\beta z) \va{e}_y$ ($\va{e}_\alpha$ is the unit vector along the $\alpha$-direction with $\alpha=\{x,y,z\}$), where the plasmon mode index $\beta$ and its field profile $H_y(x)$ are obtained by solving the wave equation given by
\begin{equation}
	\label{eq:Maxwell}
	\epsilon_{xx}(x) \frac{\dd{}}{\dd{x}} \bigg[ \frac{1}{\epsilon_{zz}(x)} \frac{\dd{H_y(x)}}{\dd{x}} \bigg] + k_0^2 \epsilon_{xx}(x) H_y(x) = \beta^2 H_y(x) \, . 
\end{equation}
Here, $\epsilon_{zz}(x)$ and $\epsilon_{xx}(x)$ are the parallel and perpendicular parts of the position-dependent dielectric function of the waveguide, respectively, and $k_0=\omega/c$ is the vacuum wavevector. In the present work, we neglect the anisotropy of the dielectric function for simplicity, and assume that $\epsilon_{xx}(x)=\epsilon_{zz}(x)=\epsilon(x)$. The anisotropy of the dielectric function has a negligible effect on the plasmon mode indices of the slab \cite{Enok2019}, but it can be added straightforwardly if required. For a general dielectric profile, this eigenvalue problem should be solved numerically. Note that the electric field can be readily obtained from the magnetic field as $\va{E}=(E_x\va{e}_x+E_z\va{e}_z)\exp(i\beta z)$ with $E_z(x)=i[\omega\epsilon_0\epsilon(x)]^{-1}\dd{H_y}/\dd{x}$ and $E_x(x)=[\omega\epsilon_0\epsilon(x)]^{-1}\beta H_y(x)$. In addition, the $z$-component of the complex poynting vector is given by $S_z(x,z) \equiv2^{-1}(\va{E}\times\va{H}^*)\cdot\va{e}_z=2^{-1} E_x(x)H_y^*(x) \exp(-2\beta_iz) = [2\omega\epsilon_0\epsilon(x)]^{-1}\beta |H_y(x)|^2 \exp(-2\beta_iz)$, in which the propagation loss (related to the real part of $S_z$) originates from the imaginary part of the mode index, $\beta_i$.

\subsection{Classical solution \label{sec:Classical}}
In the classical case, where the QSO is neglected, the interfaces between different material regions are modeled by sharp boundaries. The dielectric function of the slab waveguide then reads $\epsilon^{(0)}(x) \equiv \epsilon_1+(\epsilon_m-\epsilon_1) \theta(d/2-|x|)+(\epsilon_2-\epsilon_1) \theta(x-d/2)$. Here, $\theta$ denotes the Heaviside step function, $d$ is the slab thickness, and $\epsilon_m$, $\epsilon_1$ and $\epsilon_2$ are the bulk metal, superstrate and substrate dielectric constants, respectively. Note that we label all classical variables by superscript "0" or subscript "0". Also, we use the experimental values for the dielectric constant of bulk gold reported in Ref.~\cite{Johnson1972} for $\epsilon_m$. For $\epsilon^{(0)}(x)$, Eq.~(\ref{eq:Maxwell}) can be solved analytically by finding the solutions in three regions, \ie slab, substrate and superstrate, and matching these solutions across the two interfaces, \ie requiring $H_y(x)$ and $E_z(x)\propto\epsilon^{-1}(x)\dd{H_y}/\dd{x}$ to be continuous functions. Therefore, the bound modes are given by
\begin{equation}
	H_y^{(0)}(x) = \left\{ \begin{array}{lr}
	A_1 e^{\kappa_1 x} \qquad \qquad & x<-d/2 \\
	B_1 \cos(q x) + B_2 \sin(q x) \qquad \qquad & |x|<d/2 \\
	A_2 e^{-\kappa_2 x} \qquad \qquad & x>d/2 \, , \\
	\end{array} \right.
\end{equation}  
where $q \equiv \sqrt{\epsilon_mk_0^2-\beta_0^2}$ and $\kappa_i \equiv \sqrt{\beta_0^2-\epsilon_i k_0^2}$. Matching the partial solutions across the interfaces, the dispersion relation is obtained as $q\epsilon_m(\epsilon_1 \kappa_1+\epsilon_2\kappa_2)=(\epsilon_1\epsilon_2q^2-\epsilon_m^2\kappa_1\kappa_2) \tan(qd)$. In a symmetric geometry, where $\epsilon_1=\epsilon_2$, two modes with even ($A_1=A_2$ and $B_2=0$) and odd ($A_1=-A_2$ and $B_1=0$) symmetries are formed. The dispersion relations for the even and odd modes read $\epsilon_m\kappa=\epsilon_1q\tan(qd/2)$ and $\epsilon_m \kappa=-\epsilon_1q\cot(qd/2)$, respectively. Even (odd) modes are typically referred to as long-range (short-range) SPPs, since the imaginary parts of their mode indices are smaller (larger) than the single metal-dielectric interface SPPs. Hereafter, we only consider symmetric geometries for simplicity, but our results can be readily generalized to include non-symmetric structures. Furthermore, without loss of generality, we set $A_1=\exp(\kappa_rd/2)$ with $\kappa_r$ the real part of $\kappa \equiv \kappa_1=\kappa_2$, which leads to $|H_y^{(0)}(\pm d/2)|=1$. Note that although determining $\beta_0$ from the dispersion relation involves finding roots numerically, we refer to the classical solution as the analytical one, since this numerical task is straightforward.


\subsection{Numerical solution: Fourier modal method}
The FMM is a variant of the modal methods, in which the structure is discretized into layers and the eigenmodes of each layer are connected by mode matching at the interfaces \cite{Lavrinenko2018}. In the FMM, which is also referred to as rigorous coupled wave analysis (RCWA), the eigenmodes are expanded using a Fourier basis.
The FMM has been used widely for investigating gratings \cite{Moharam1995, Granet1996, Taghizadeh2015, Yoon2015, Learkthanakhachon2016, Orta2016}, since the Fourier basis is particularly suitable for periodic structures. Nonetheless, by introducing perfectly matched layers (PMLs) at the boundaries of simulation domains, the FMM has been successfully employed for studying various non-periodic structures such as dielectric waveguides \cite{Hugonin2005, Silberstein2001, Ctyroky2010}, photonic crystals waveguides \cite{Pissoort2004, Lecamp2007}, finite gratings \cite{Pisarenco2010}, microdisks \cite{Armaroli2008, Bigourdan2014}, and vertical cavities \cite{Taghizadeh2016, Taghizadeh2017}. For the waveguide problems, the PML is typically implemented as a nonlinear coordinate transformation \cite{Hugonin2005}, in which the infinite space $x$ is mapped to a finite space $x'$ with a width of $\Lambda$, \ie $x=F(x')$. In the transformed space, the Fourier basis is then used for expanding the field $H_y(x')=\sum_n h_n \exp(i k_n x')$, dielectric function $\epsilon(x')=\sum_n \varepsilon_n \exp(i k_n x')$, its inverse $\epsilon^{-1}(x')=\sum_n \eta_n \exp(i k_n x')$ and the derivative of the coordinate transformation $f(x')=\dd{F}/\dd{x'}=\sum_n f_n \exp(i k_n x')$. Here, $k_n \equiv 2\pi n/\Lambda$ and the summations over $n$ run from $-N$ to $N$ with a total number of $N_t=2N+1$ basis functions.

Using the Fourier expansions, Eq.~(\ref{eq:Maxwell}) will be transformed to a matrix eigenvalue problem given by \cite{Taghizadeh2016}
\begin{equation}
\label{eq:FMM}
	\va{A}^{-1}(k_0^2\va{I}-\va{F}\va{K} \va{E}^{-1} \va{F} \va{K}) \va{h} = \beta^2 \va{h}
\end{equation}
where $\va{K}=[[k_n\delta_{nm}]]$ is a diagonal matrix with entries $k_n$ ($\delta_{nm}$ is the Kronecker delta), and $\va{h}=[h_n]$ is a vector formed by $h_n$. In addition, $\va{A}=[[\eta_{n-m}]]$, $\va{E}=[[\varepsilon_{n-m}]]$ and $\va{F}=[[f_{n-m}]]$ are the Toeplitz matrices associated with $\eta_n$, $\varepsilon_n$ and $f_n$, respectively. Note that we have implemented the FMM with special care for TM polarization, in order to obtain rapidly converging results as discussed in Refs.~\cite{Li1996, Granet1996}. This is due to the fact that the Fourier coefficients of products of two functions having jump discontinuities converge much faster if the inverse rule is applied for obtaining the Fourier coefficients \cite{Li1996}.
In addition, we have modified the coordinate transformation suggested in Ref.~\cite{Hugonin2005} as explained in Appendix~\ref{sec:AppC}, in order to improve the convergence of our numerical results for the long-range SPP.

\subsection{Perturbative solution}
Perturbation methods have been highly successful in quantum mechanics for investigating the effects of small modifications on various quantum systems \cite{Griffiths2005}. They provide not only effective numerical tools for computations but also means of gaining useful physical insight into the systems. Surprisingly, their use has been restricted in optics, presumably due to the vectorial nature of Maxwell's equations and the boundary conditions at the interfaces \cite{Johnson2002}. Nonetheless, here, we formulate a perturbative approach for obtaining the effect of QSO on the plasmon mode indices. This is done by noticing that the magnetic field $H_y$ and transverse component of the electric field $E_z$ of the plasmon modes are only modified slightly, when the QSO is included, compared to the case without QSO, \ie $H_y(x) \approx H_y^{(0)}(x)$ and $\epsilon^{-1}(x)\dd{H_y}/\dd{x} \approx [\epsilon^{(0)}(x)]^{-1}\dd{H_y^{(0)}}/\dd{x}$ as shown in Appendix~\ref{sec:AppA}. Subsequently, we consider the integral of the complex poynting vector across the waveguide (the complex energy flux) at $z=0$, \ie $P=\beta(2\omega\epsilon_0)^{-1} \int \dd{x} |H_y(x)|^2[\epsilon(x)]^{-1}$. One can readily show that $\beta P \approx \beta_0 P^{(0)}$, since 
\begin{equation}
	 P \propto \beta\int \dd{x} \frac{|H_y|^2}{\epsilon(x)} = \int \dd{x} \frac{H_y^*}{\beta}  \Bigg\{ \frac{\dd{}}{\dd{x}}\bigg[ \frac{1}{\epsilon(x)} \frac{\dd{H_y}}{\dd{x}} \bigg] + k_0^2 H_y(x) \Bigg\} \approx \frac{\beta_0^2}{\beta} \int \dd{x} \frac{|H_y^{(0)}|^2}{\epsilon^{(0)}(x)} \propto \dfrac{\beta_0}{\beta} P^{(0)} \, . 
\end{equation} 
Here, the pre-factor $(2\omega\epsilon_0)^{-1}$ is omitted in the first and last relations, and Eq.~(\ref{eq:Maxwell}) has been used in the second and third relations. Now, by replacing $H_y$ by $H_y^{(0)}$ in the original expression for $P$, the following simple approximation is derived for the plasmon mode index:
\begin{equation}
	\label{eq:Perturbation}
	\beta^2 \approx \beta_0^2 \dfrac{\int \dd{x} |H_y^{(0)}(x)|^2[\epsilon^{(0)}(x)]^{-1} }{\int \dd{x} |H_y^{(0)}(x)|^2\epsilon^{-1}(x) } = \beta_0^2 \dfrac{t_0}{t_0+\Delta t} \, .
\end{equation}
Here, $t_0$ and $\Delta t$ are defined as $t_0 \equiv \int \dd{x} |H_y^{(0)}(x)|^2[\epsilon^{(0)}(x)]^{-1}$ and $\Delta t \equiv \int \dd{x} |H_y^{(0)}(x)|^2[1/\epsilon(x)-1/\epsilon^{(0)}(x)]$, respectively. As shown in Appendix~\ref{sec:AppB}, $t_0$ can be calculated analytically, whereas $\Delta t$ should be computed numerically for a general dielectric profile $\epsilon(x)$. Note that the integrand in $\Delta t$ is only non-negligible in the vicinity of the interfaces, which makes its computation straightforward. It is expected that $|\Delta t| \ll |t_0|$ when the QSO is a small perturbation, and hence $\beta \approx \beta_0 (1-\Delta t/2t_0)$. Equation~(\ref{eq:Perturbation}) can be readily used to compute the effect of QSO on the mode indices with a impressive accuracy as shown in the following section.  


\section{Numerical results \label{sec:Numerics}}
In this section, we study the performance of the FMM and perturbative approach for computing the plasmon mode indices of gold slabs. The dielectric function of the gold slab with QSO is calculated using DFT based on a jellium model as explained in details in Ref.~\cite{Enok2019}. The jellium model provides the position-dependent electron density $n(x)$ due to the free electrons, which converges to the bulk electron density $n_0$ near the slab midpoint. The effect of bound electrons in the lower $d$-bands is added to retain the experimental value of the bulk dielectric constant of gold in the central part of the slab. Hence, the contribution of bound electrons in the dielectric constant is assumed to be $\epsilon_b=\epsilon_m-\epsilon_D$, where $\epsilon_D=1-\omega_{p0}^2/(\omega^2+i\omega\gamma)$ is the Drude response of the bulk gold ($\omega_{p0}=\sqrt{n_0e^2/(m\epsilon_0)} \approx 9.025$ eV and $\gamma=65.8$ meV \cite{Novotny2012}). 
Therefore, the position-dependent dielectric function $\epsilon(x)$ of the gold slab reads
\begin{equation}
	\epsilon(x) = 1-\frac{\omega_p^2(x)}{\omega(\omega+i \gamma)} + \epsilon_b \theta(d/2-|x|) + (\epsilon_1-1) \theta(|x|-d/2) \, , 
\end{equation}
where $\omega_p(x)=\sqrt{n(x)e^2/(m\epsilon_0)}$. A typical calculated dielectric function for a 2 nm gold slab surrounded by glass (on both sides) is shown in Fig.~\ref{fig:Schematic}(b). Note that the abrupt jumps ($\Delta\epsilon=\epsilon_b-\epsilon_1+1$) at two interfaces originate mainly from the bound electron term.

\begin{figure}[t]
	\centering\includegraphics[width=0.8\textwidth]{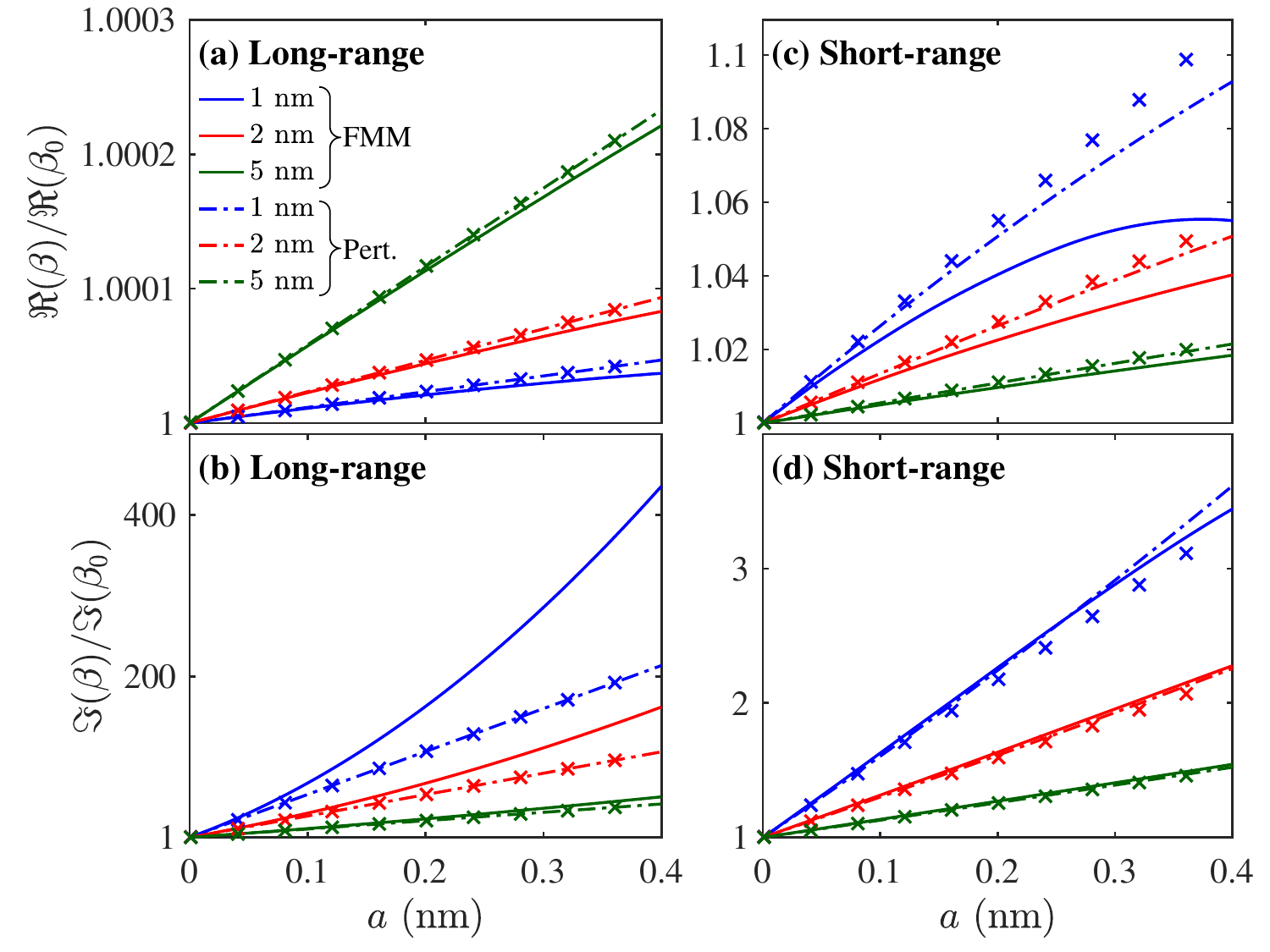}
	\caption{Real (a,c) and imaginary (b,d) parts of the long-range (a,b) and short-range (c,d) SPP mode indices $\beta$ with QSO, normalized to the corresponding values without QSO, $\beta_0$, versus the spill-out parameter $a$. The calculations are performed for three waveguide widths: $d=1$ nm (blue), $d=2$ nm (red), $d=5$ nm (green). The results are obtained for a gold slab ($\epsilon_m=-21.995+ 1.363i$\cite{Johnson1972} and $\epsilon_b=8.778 + 0.056i$) surrounded by glass ($\epsilon_1=2.25$) at the wavelength of 775 nm, modeled by the dielectric function in Eq.(\ref{eq:Tanh}). The solid-lines are obtained from the FMM by solving Eq.~(\ref{eq:FMM}) numerically, while the dashed-lines are the perturbation results using Eq.~(\ref{eq:Perturbation}). The stars show the analytical results from Eqs.~(\ref{eq:Analytic0}) and (\ref{eq:Analytic}). For small values of $a$, the effect of QSO vanishes and hence, all curves converge toward one as expected, whereas for large values of $a$ the perturbative results deviate from the numerical solutions.
		\label{fig:Pert_Limit}}
\end{figure} 

Using the DFT model, we have computed the dielectric function for various slab thicknesses, ranging from 0.3 nm to 200 nm. For various slab thicknesses, the dielectric function shows a similar pattern: a smooth variation of the response in $\sim0.5$ nm-width regions near the boundaries followed by multiple Friedel oscillations inside the metal close to the interfaces \cite{Enok2019}. After careful consideration, we conclude that the effect of QSO can be captured fairly accurately by using a toy dielectric function, $\epsilon_t(x)$, which can emulate the smooth behavior of the Drude part of the dielectric function at the interface (but not Friedel oscillations). This toy model can be used for quantitative understanding of the effect of QSO on plasmon mode indices and allows us investigate the performance of the perturbative approach systematically. Hence, we choose $\epsilon_t(x)$ to be
\begin{align}
	\label{eq:Tanh}
	\epsilon_t(x) &= 1+\frac{\epsilon_D-1}{4}(1-e^{-2d/a}) \bigg[ \tanh\Big(\frac{x+d/2}{a}\Big)+1 \bigg] \bigg[ \tanh\Big(\frac{-x+d/2}{a}\Big)+1 \bigg] \nonumber \\
	& \hspace{5cm} + \epsilon_b \theta(d/2-|x|) + (\epsilon_1-1) \theta(|x|-d/2) \, ,
\end{align}
where $a$ is a spill-out parameter with dimensions of length, that captures the degree of QSO. Also, $1-\exp(-2d/a)$ is introduced in order to conserve the average of the dielectric function, \ie $\int \epsilon_t(x)\dd{x}=\int \epsilon^{(0)}(x)\dd{x}$. In Fig.~\ref{fig:Schematic}(b), we compare $\epsilon_t(x)$ with the DFT calculated dielectric function for a gold slab surrounded by glass, where the spill-out parameter is fitted to be $a=0.09$ nm. The QSO affects the dielectric function mainly close to the interfaces in a region that depends on $a$ (approximately $d/2-3a<|x|<d/2+3a$).
The toy dielectric function will converge toward the classical case by letting $a\rightarrow 0$, and the strength of the perturbation can be tuned by varying $a$. Furthermore, we can compute $\Delta t$ analytically as discussed in Appendix~\ref{sec:AppB}, and derive a closed form expression for $\beta$. This expression provides a simple measure of the impact of QSO on plasmon mode indices.

\begin{figure}[t]
	\centering\includegraphics[width=0.8\textwidth]{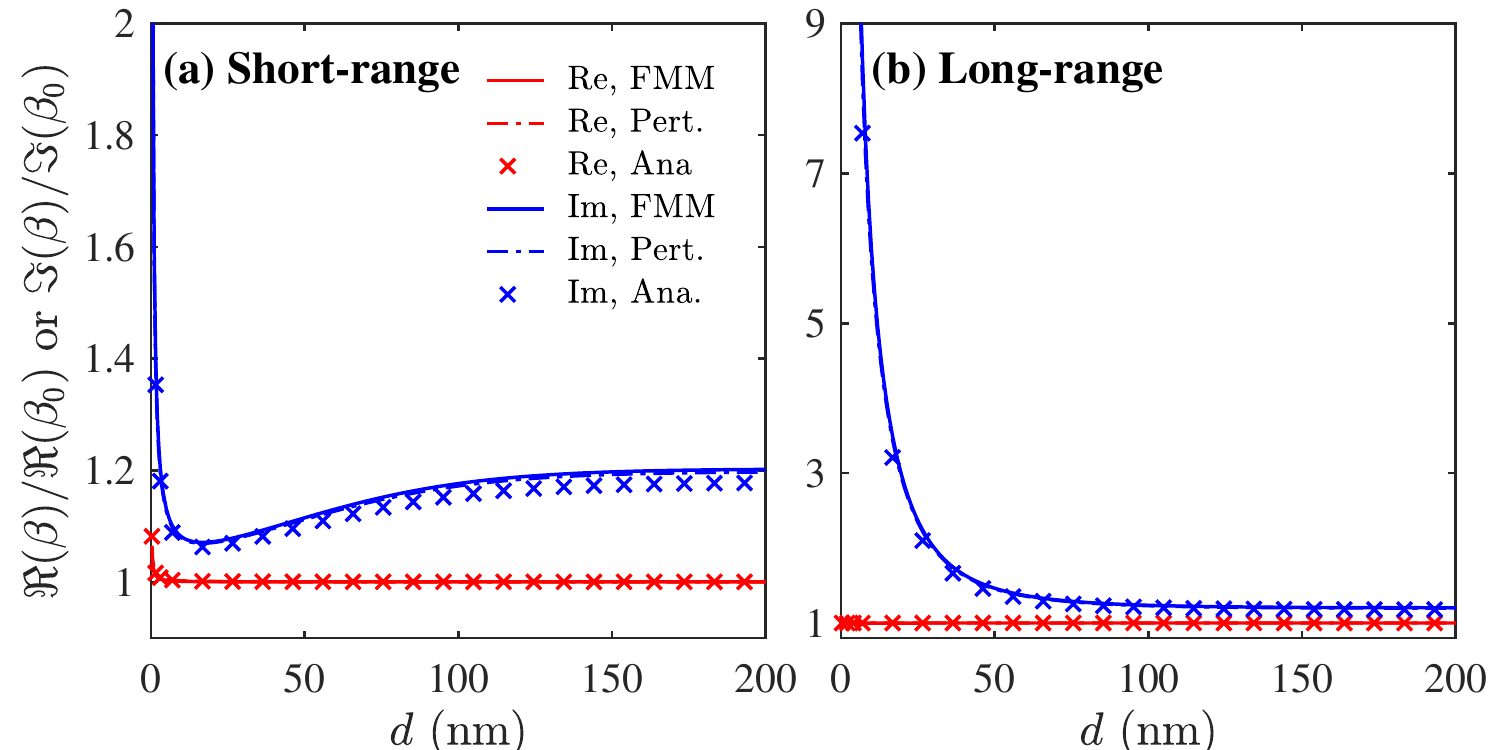}
	\caption{Real (red) and imaginary (blue) parts of the short-range (a) and long-range (b) SPPs with QSO, $\beta$, normalized to the corresponding values without QSO, $\beta_0$, versus the waveguide width $d$. Three different methods are used: numerical method (FMM), perturbation approach with the DFT-based dielectric profile (Pert.) and analytical solution (Ana.) The gold slab is surrounded by glass and the results are obtained at 775 nm wavelength, see caption of Fig.~\ref{fig:Pert_Limit}.
	\label{fig:Changed}}
\end{figure}

We compute both real and imaginary parts of the short/long-range mode indices, $\beta$, for the toy dielectric function and normalize them to those of the corresponding classical value, $\beta_0$, and plot them versus the spill-out parameter $a$ for three different slab widths $d=\{1,2,5\}$ nm in Fig.~\ref{fig:Pert_Limit}. Three different approaches are employed for each set of calculations: (i) FMM by solving Eq.~(\ref{eq:FMM}) numerically, (ii) the perturbative approach by numerical evaluation of Eq.~(\ref{eq:Perturbation}), and (iii) the analytical approach by employing  Eqs.~(\ref{eq:Analytic0}) and (\ref{eq:Analytic}). Since the effect of QSO vanishes if $a \rightarrow 0$, all graphs approach one for small values of $a$. 
The results show that the perturbative solutions, either using the analytical expressions or numerical integrations, are remarkably accurate for $a<0.1$ nm in all cases. Since the value of $a$ is smaller than 0.1 nm for gold, one can expect that the perturbative approach performs very well for computing the effect of QSO also if the more realistic dielectric profile based on DFT is used as demonstrated below. Furthermore, the analytical model works as well as the numerical perturbation approach, which makes it very useful for predicting the dependence of mode indices on various parameters of the problem. For instance, the linear dependence of the perturbative results on $a$ follows directly from Eq.~(\ref{eq:Analytic}).

Now we proceed to the more realistic dielectric function obtained from DFT, and compare its results with the toy model. We keep the frequency constant and vary the slab thickness from 0.3 to 200 nm. Both real and imaginary parts of plasmon mode indices for the short- and long-range modes are normalized to the corresponding values of the classical case and plotted in Figs.~\ref{fig:Changed}(a) and \ref{fig:Changed}(b), respectively. Note that the numerical FMM results are identical to the results reported in Ref.~\cite{Enok2019}, obtained from the transfer matrix method. The real parts of the mode indices for both plasmon modes are approximately unaffected by the QSO as the ratios for the real parts are nearly one. In contrast, the imaginary parts change drastically for thin slabs due to the QSO, while it saturates to $\sim1.2$ for thick slabs. In particular, the effect of QSO is evident for very wide slabs, which proves its significant role even for commonly fabricated metallic slabs. Comparing the different methods for computing the graphs in Fig.~\ref{fig:Changed}, we find that the numerical and perturbative results are in excellent agreement. Hence, the perturbation method can be used safely for computing plasmon mode indices, without using advanced numerical techniques. Furthermore, the analytical expressions (based on the toy dielectric function) capture basically all features of the graphs accurately. For instance, we derive closed form expressions for the saturation values at large $d$ in Eqs.~(\ref{eq:InfinitedA}) and (\ref{eq:InfinitedA}), which agree fairly well with the full numerical results. Also, Eq.~(\ref{eq:InfinitedA}) shows that the QSO modifies the imaginary part of the plasmon modes by an amount that is linearly proportional to the electron density spill-out.  Hence, the analytical expressions can be used to estimate quantitatively the effect of QSO on the plasmon mode indices and avoid solving the challenging problem of DFT in combination with Maxwell's equations.

\section{Conclusion \label{sec:Conclusion}}
In summary, we have computed the plasmon mode indices of a metallic slab using the FMM, with inclusion of QSO. A perturbative approach is then proposed, which reproduces the numerical results very accurately with much less effort. Using a simple model for dielectric function with QSO, analytical expressions have been derived for the mode indices of the short- and long-range SPPs. The analytical results agree very well with the numerical solutions, and may be used to obtain more physical insight into the impact of QSO on other plasmonic structures.

\appendix
\section*{Appendices}

\section{Perturbation validity \label{sec:AppA}}
\begin{figure}[t]
	\centering\includegraphics[width=0.8\textwidth]{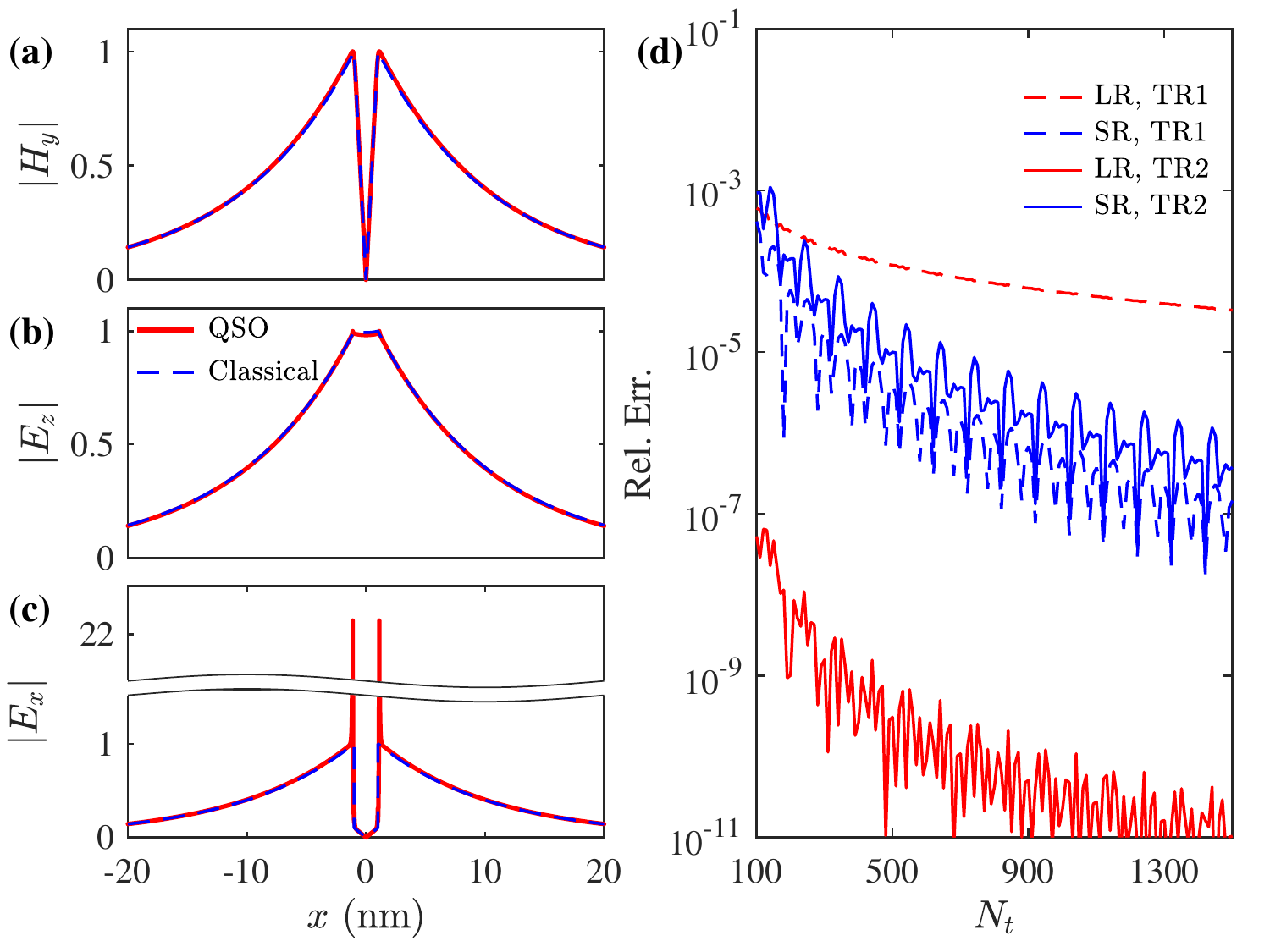}
	\caption{Normalized amplitude of (a) $H_y(x)$, (b) $E_z(x)$, (c) $E_x(x)$ profiles obtained for the short-range SPP of a 2 nm-thick gold slab, when QSO is included (red) or neglected (blue). QSO mainly modifies $E_x$ in a narrow region outside the slab and close to the interfaces where the field amplitude becomes very large. (d) The relative error of mode indices for the short-range (blue) and long-range (red) SPPs versus the number of Fourier functions, $N_t$, for two coordinate transformations: TR1 the transformation suggested in Ref.~\cite{Hugonin2005}, and TR2 the transformation in Eq.~(\ref{eq:Transformation}). The relative error is defined as $|(\beta-\beta_0)/\beta_0|$, where $\beta$ is obtained by using the FMM for the classical dielectric function, $\epsilon^{(0)}(x)$.
		\label{fig:Conv_Pert}}
\end{figure}

To demonstrate that the modifications of $H_y$ and $E_z$ due to QSO are negligible, we compare the field profiles of the short-range SPP obtained with or without QSO in Figs.~\ref{fig:Conv_Pert}(a) and \ref{fig:Conv_Pert}(b) for the 2 nm-thick gold slab of Fig.~\ref{fig:Schematic}(b). The results show that the parallel components of the fields, $H_y(x)$ and $E_z(x)$, change only slightly when QSO is included. Note that the perpendicular component of the electric field, $E_x(x)$, is drastically modified as shown in Fig.~\ref{fig:Conv_Pert}(c), which is the origin of the increased imaginary part of the plasmon mode index. The drastic change is due to the fact that the real part of the dielectric function is crossing zero at a point outside the slab, which leads to a large value for $E_x(x) \propto H_y(x)/\epsilon(x)$ as seen in Fig.~\ref{fig:Conv_Pert}(c).

\section{Analytical expressions \label{sec:AppB}}
The perturbation method allows us to derive analytical expressions, when the QSO is included, for simple geometries and some forms of dielectric function such as Eq.~(\ref{eq:Tanh}). For symmetric geometries, an analytical equation can be derived for $t_0$ as
\begin{equation}
	\label{eq:Analytic0}
	t_0 = \dfrac{1}{\epsilon_1 \kappa_r} + \dfrac{1}{\epsilon_m} \dfrac{\sinh(q_i d)/q_i \pm \sin(q_r d)/q_r}{\cosh(q_i d) \pm \cos(q_r d)} 
\end{equation}
where $r$ and $i$ subscripts denote the real and imaginary parts of the corresponding parameter, respectively, and $+$ ($-$) is used for the long-range (short-range) SPP. Regarding $\Delta t$, we can derive an analytical expression employing $\epsilon_t(x)$ in Eq.~(\ref{eq:Tanh}) by noticing that the integrand in $\Delta t$ is only non-zero for a narrow region in the vicinity of $|x|=d/2$. Moreover, the field $H_y^{(0)}$ does not vary considerably in this narrow region, and hence  
\begin{align}
	\label{eq:Analytic}
	\Delta t \approx & 2|H_y^{(0)}(d/2)|^2 \int_{d/2}^{\infty} \dd{x} \bigg\{ \dfrac{2}{\epsilon_p+2\epsilon_1+\epsilon_p\tanh[(d/2-x)/a]}-\dfrac{1}{\epsilon_1} \bigg\} \nonumber \\ 
	+&2|H_y^{(0)}(d/2)|^2 \int_{-\infty}^{d/2} \dd{x} \bigg\{ \dfrac{2}{\epsilon_m+\epsilon_b+1+\epsilon_p\tanh[(d/2-x)/a]}-\dfrac{1}{\epsilon_m} \bigg\} \nonumber \\ 
	=& a\epsilon_p \bigg[ \dfrac{ \log(2-\epsilon_p/\epsilon_m)}{\epsilon_m(\epsilon_m-\epsilon_p)} - \dfrac{\log(2+\epsilon_p/\epsilon_1)}{\epsilon_1(\epsilon_1+\epsilon_p)}  \bigg] \, ,
\end{align}
Here, $\epsilon_p\equiv\epsilon_D-1$ and it is assumed that $d/a \gg 1$, which simplifies the expression for $\epsilon_t$ by letting $\exp(-2d/a) \rightarrow 0$ and $\tanh[(d/2+x)/a] \rightarrow 1$.  
Equations~(\ref{eq:Analytic0}) and (\ref{eq:Analytic}) can be employed to derive useful expressions for the saturation levels in Fig.~\ref{fig:Changed} by taking the limit $d\rightarrow\infty$ as
\begin{subequations}
	\begin{align}
	\label{eq:InfinitedA}
	&\dfrac{\Re(\beta)}{\Re(\beta_0)} \bigg|_{d\rightarrow\infty} \approx 1+\dfrac{ak_0\epsilon_1}{2\sqrt{-\epsilon_{mr}}}\Re[\log(2+\epsilon_p/\epsilon_1)] \, , \\
	\label{eq:InfinitedB}
	&\dfrac{\Im(\beta)}{\Im(\beta_0)} \bigg|_{d\rightarrow\infty} \approx \dfrac{\Re(\beta)}{\Re(\beta_0)}\bigg|_{d\rightarrow\infty} + \bigg(\dfrac{2\epsilon_{mr}^2}{\epsilon_1\epsilon_{mi}}\bigg) \dfrac{ak_0\epsilon_1}{2\sqrt{-\epsilon_{mr}}}\Im[\log(2+\epsilon_p/\epsilon_1)] \, .
	\end{align}
\end{subequations}
Here, $\epsilon_{mr}\equiv\Re(\epsilon_m)$, $\epsilon_{mi}\equiv\Im(\epsilon_m)$, and we assume that $|\epsilon_m|,|\epsilon_D| \gg \epsilon_1, \epsilon_{mi}$. Equations~(\ref{eq:InfinitedA}) and (\ref{eq:InfinitedB}) show that the saturation levels will increase linearly with $a$, and confirm the non-negligible modification of the imaginary ratio due to the appearance of the $2\epsilon_{mr}^2/(\epsilon_1\epsilon_{mi})$ factor, which can be large. These expressions predict the saturation values to be 1.0004 and 1.171 for the real and imaginary parts, respectively, which are in reasonable agreement with the results in Fig.~\ref{fig:Changed}. 

\section{Coordinate transformation for Fourier modal method \label{sec:AppC}}
A nonlinear coordinate transformation is suggested in Ref.~\cite{Hugonin2005} (referred to as TR1 hereafter), which typically works well for dielectric waveguide structures. However, TR1 is not sufficiently accurate for the plasmonic waveguides in this work, in particular for long-range SPPs, since the plasmon modes decay very slowly in the dielectric regions. Therefore, we modify TR1 by introducing a second mapped region as shown in Fig.~\ref{fig:Schematic}(a), which leads to improvement of the convergence rate. The transformation of the infinite space $x$ to the finite space $x'$ with the length $\Lambda$ is defined as
\begin{equation}
\label{eq:Transformation}
	x = F(x') = \left\{ \begin{array}{ll}
	x' & |x'| \le d_1/2 \\
	\dfrac{x'}{|x'|}  \dfrac{d_1}{2} e^{2|x'|/d_1-1}  & d_1/2 < |x'| \le d_2/2 \\
	\dfrac{x'}{|x'|} e^{2d_2/d_1-1} \Bigg[ \dfrac{d_1}{2}+\dfrac{\Lambda-d_2}{2} \tan\left(\pi\dfrac{|x'|-d_2/2}{\Lambda-d_2}\right) \Bigg] \qquad & d_2/2 < |x'| \le \Lambda/2 \, .  \\
	\end{array} \right.
\end{equation}
This transformation and its derivative are continuous functions, and the Fourier coefficients of its derivative $f(x')\equiv\dd{F}/\dd{x'}$ [which are required in Eq.~(\ref{eq:FMM})] can be obtained analytically. We refer to this transformation as TR2. There are three length parameters in the TR2 $\{d_1, d_2, \Lambda \}$, which are chosen to achieve the required convergence. In particular, we choose $d_1$ and $d_2$ such that $\exp(d_2/d_1-1)$ is larger than the mode extension inside the dielectric regions. Note that by setting $d_1=d_2$, TR1 can be recovered. To study the performance of the coordinate transformation, we compare the mode indices obtained for the short- and long-range SPPs of a gold slab (without QSO) for the two transformations (TR1 and TR2) as a function of $N_t$ in Fig.~\ref{fig:Conv_Pert}(d). The relative error is defined with respect to the exact solution of the metallic slab, see \sref{sec:Classical}. The results show that using TR2 improves the convergence rate for the long-range SPP. Note that TR2 does not change the convergence performance of the short-range SPP in this example, but it can improve the convergence rate for thicker slabs.

\section*{Funding}
Danish National Research Foundation (Project No. DNRF103).


\section*{Disclosures}
The authors declare no conflicts of interest.


\bibliography{Spillout}

\end{document}